\documentclass[fleqn,11pt,twoside]{article}

\usepackage{amsthm,amsthm,amssymb, color, xcolor,epsfig, graphics, subfigure}

\usepackage{amsmath, graphicx, latexsym, lscape }

\makeatletter
\newcommand{\copyrightnote}[2]{{\renewcommand{\thefootnote}{}
 \footnotetext{\small\it
\begin{flushleft}
 \copyright \ #1   #2  
\end{flushleft}}}}

\newcommand{\Name}[1]{\begin{flushleft}
                       \LARGE \bf #1
                       \end{flushleft}\vspace{-3mm}}

\newcommand{\Author}[1]{\begin{flushleft}
                       \it #1 \end{flushleft}}

\newcommand{\Address}[1]{\begin{flushleft}
                       \it #1 \end{flushleft}}

\newcommand{\Date}[1]{\begin{flushleft}
                      \small  \it #1 \end{flushleft}}

\newcommand{\evenhead}{Author \ name}
\newcommand{\oddhead}{Article \ name}

\renewcommand{\@evenhead}{
\hspace*{-3pt}\raisebox{-15pt}[\headheight][0pt]{\vbox{\hbox to \textwidth
{\thepage \hfil \evenhead}\vskip4pt \hrule}}}
\renewcommand{\@oddhead}{
\hspace*{-3pt}\raisebox{-15pt}[\headheight][0pt]{\vbox{\hbox to \textwidth
{\oddhead \hfil \thepage}\vskip4pt\hrule}}}
\renewcommand{\@evenfoot}{}
\renewcommand{\@oddfoot}{}

\long\def\@makecaption#1#2{%
  \vskip\abovecaptionskip
  \sbox\@tempboxa{\small \textbf{#1.}\ \ #2}%
  \ifdim \wd\@tempboxa >\hsize
    {\small \textbf{#1.}\ \ #2}\par
  \else
    \global \@minipagefalse
    \hb@xt@\hsize{\hfil\box\@tempboxa\hfil}%
  \fi
  \vskip\belowcaptionskip}

\newcommand{\JNMPnumberwithin}[3][\arabic]{%
  \@ifundefined{c@#2}{\@nocounterr{#2}}{%
    \@ifundefined{c@#3}{\@nocnterr{#3}}{%
      \@addtoreset{#2}{#3}%
      \@xp\xdef\csname the#2\endcsname{%
        \@xp\@nx\csname the#3\endcsname .\@nx#1{#2}}}}%
}

\newcommand{\resetfootnoterule} {
  \renewcommand\footnoterule{%
  \kern-3\p@
  \hrule\@width.4\columnwidth
  \kern2.6\p@}
}

\renewcommand{\footnoterule}{}

\makeatother

\theoremstyle{definition}
\newtheorem*{definition}{Definition}
\newtheorem*{example}{Example} 

\theoremstyle{plain}
\theoremstyle{definition}
\newtheorem{theorem}{Theorem}[section]
\newtheorem{lemma}[theorem]{Lemma}
\newtheorem{definition-theorem}[theorem]{Definition-Theorem}
\newtheorem{definition-proposition}[theorem]{Definition-Proposition}

\newtheorem{proposition}[theorem]{Proposition}
\newtheorem{corollary}[theorem]{Corollary}
\newtheorem{examples}{Example}[subsection]

\newtheorem{remark}{Remark}[section]
\newtheorem{remarks}{Remarks}[section]

\numberwithin{equation}{section} 

\DeclareMathOperator{\diag}{diag}

\def\tr{\mathrm {tr}}

\def\diag{\mathrm {diag}}

\def\res{\mathop{\mathrm {res}}\limits}

\def\res{\mathop{\mathrm{res}}\limits}

\def\&{&{\hskip -20pt}}

\def\be{\begin{eqnarray}}
\def\ee{\end{eqnarray}}

\def\bt{\begin{theorem}}
\def\et{\end{theorem}}

\def\bp{\begin{proposition}}
\def\ep{\end{proposition}}

\def\bex{\begin{example}\small \rm}
\def\eex{\end{example}}

\def\bexs{\begin{examples}\small \rm}
\def\eexs{\end{examples}}

\def\br{\begin{remark}\small \rm}
\def\er{\end{remark}}

\def\Jb{\mathbf{J}}

\def\pb{\mathbf{p}}

\def\tb{\mathbf{t}}

\def\zb{\mathbf{z}}

\def\W{\hat{W}_{1+\infty}}

\newcommand{\Pa}{\mathop\mathrm{P}\nolimits}

\def\bp{\begin{Proposition}\rm}
\def\ep{\end{Proposition}}
\def\bc{\begin{corollary}}
\def\ec{\end{corollary}}
\def\bl{\begin{lemma}\em}
\def\el{\end{lemma}}
\def\br{\begin{remark}\rm\small}
\def\er{\end{remark}}
\def\brs{\begin{remarks}.\\ \rm\
\begin{enumerate}}
\def\ers{\end{enumerate}\end{remarks}}
\def\bea{\begin{eqnarray}}
\def\eea{\end{eqnarray}}

\begin{document}

\renewcommand{\evenhead}{ {\LARGE\textcolor{blue!10!black!40!green}{{\sf \ \ \ ]ocnmp[}}}\strut\hfill  A. Yu. Orlov}
\renewcommand{\oddhead}{ {\LARGE\textcolor{blue!10!black!40!green}{{\sf ]ocnmp[}}}\ \ \ \ \  
$W_{1+\infty}$ flows and multi-component hierarchy. KP case}

\thispagestyle{empty}
\newcommand{\FistPageHead}[3]{
\begin{flushleft}
\raisebox{8mm}[0pt][0pt]
{\footnotesize \sf
\parbox{150mm}{{Open Communications in Nonlinear Mathematical Physics}\ \  {\LARGE\textcolor{blue!10!black!40!green}{]ocnmp[}}
\ Vol.4 (2024) pp
#2\hfill {\sc #3}}}\vspace{-13mm}
\end{flushleft}}

\FistPageHead{1}{\pageref{firstpage}--\pageref{lastpage}}{ \ \ Article}

\strut\hfill

\strut\hfill

\copyrightnote{The author(s). Distributed under a Creative Commons Attribution 4.0 International License}

\Name{$W_{1+\infty}$ flows and multi-component hierarchy. \\KP case}

\Author{A. Yu. Orlov }

\Address{Institute of Oceanology, Nahimovskii Prospekt 36, Moscow 117997, Russia;\\
National Research University Higher School of Economics\\[0.3cm]
Email: orlovs@ocean.ru}

\Date{Received February 14, 2024; Accepted August 14, 2024}

\setcounter{equation}{0}

\begin{abstract}
\noindent 
We show that abelian subalgebras of generalized $W_{1+\infty}$ ($GW_{1+\infty}$) 
algebra gives rise to the multicomponent KP flows. The matrix elements of the related group elements in the fermionic Fock space is expressed as a product 
of a certain factor (generalized content product) and
of a number of the Schur functions and the skew Schur functions.

\end{abstract}

\label{firstpage}


\section{Introduction}

This work is the remark concerning a question addressed in very old paper \cite{Orlov1988}
about commuting subhierarchies of the ``additional symmetries'', 
later known as $\hat{W}_{1+\infty}$. 

In KP theory there is the current algebra whose abelien part produce what is called
KP higher flows.

I want to choose any of symmetry operator from $\W$ of a given degree and construct a
hierarchy of operators commuting with chosen one, construct multiparametric group
flows and present the answer.

The example is as follows. We choose the Virasoro algebra element $L_{-1}\in W_{1+\infty}$ written in the KP hierarchy higher times and construct the graded abelian algebra
which contains $L_{-1}$. Then the related abelian flows gives rise
 the known \cite{OS},\cite{HO2003} expression for the two-matrix model:
\be\label{2MM}
\int e^{\sum_{n>0} 
\left(\frac{p_n}{n}\tr\left( X\right)^n+\frac{p^*_n}{n}\tr\left( Y\right)^n \right)  + 
\sqrt{-1}\tr\left( XY \right)}   
\prod_{i\ge j} d\Re X_{i,j}d\Re Y_{i,j}\prod_{i<j} d\Im X_{i,j}d\Im Y_{i,j}\nonumber\\
=
e^{\sum_{n>0}\frac{p^*_n}{n} L^{(n)}(\pb,N)}\cdot 1 ,
\ee
where the exponential in the right-hand side is the multiparameter group action of the 
abelian flows generated by operators $L^{(n)}\in W_{1+\infty}$, $n=1,2,\dots$,
\be
[L^{(n)},L^{(m)}]=0,\quad n,m =1,2,\dots,
\ee
where
\be
L^{(1)}(\pb,N)=N p_1 +\sum_{i>0} \frac1i p_{i+1}\frac{\partial}{\partial p_i}.
\ee
In the right-hand side, $p^*_n$ are the group parameters, which are also coupling constants of the matrix model on the left-hand side of (\ref{2MM}).
Then, for the one-matrix model one obtains
\be
\int e^{\frac14\tr\left( X \right)^2+\sum_{n>0} \frac{p_n}{n}\tr\left( X\right)^n  }  
\prod_{i\ge j} d\Re X_{i,j}\prod_{i<j} d\Im X_{i,j}=e^{L^{(2)}(\pb,N)}\cdot 1
\ee

If we assign $\deg p_n = n$, $\deg p^*_n = -n$ then we are looking for commuting generators  $L^{(n)}$,
where $\deg L^{(n)}=n$   and $L^{(1)}$ is given. We call such sets of graded elements
abelian hierarchy.

More general example is the so-called hypergeometric tau function:
\be\label{hyptau}
\tau^{\rm hyp}(\pb,\pb^*)=\sum_\lambda r_\lambda s_\lambda(\pb)s_\lambda(\pb^*),
\ee
which can be also obtained as the action of certain abelian group of $W_{1+\infty}$ algebra,
where $p^*_n$ are group parameters; see \cite{OS}. In (\ref{hyptau}) $s_\lambda$ is the 
Schur polynomial, $\deg s_\lambda =|\lambda|$, where $\lambda$ is labeled by a partition
$\lambda=(\lambda_1,\dots,\lambda_N)$, $|\lambda|=\lambda_1+\cdot +N $ is the weight of
$\lambda$ (see Appendix) and
\be\label{contentProd}
r_\lambda = \prod_{(i,j)}r(j-i)
\ee
is the so-called (generalized) content product, given by the choice
$\{r(i)=r_i,\, i \in Z\}$ which can be vied either as a function of the lattice $Z$,  or just as a set of numbers.

We will use the fermionic approach because it is much more compact than any other.

We will generalize the construction given in \cite{OS}.
In fact, in the next section we shall consider the abelian subalgebra of ${\widehat {gl}}_\infty$ generated by 
graded elements (we call it the abelian hierarchy).

\section{$\widehat{gl}_\infty$ algebra, $\W$ algebra and fermions.}

Let us remind some notions and facts known from the seminal
works of Kyoto school; see \cite{JM} and references therein.

The modes $\psi_i,\psi^\dag_i$ of the Fermi operators on the cirle satisfy
the canonical relations
\be\label{canon}
[\psi_i,\psi_j]_+=0=[\psi^\dag_i,\psi^\dag_j]_+,\quad
[\psi_i,\psi^\dag_j]_+=\delta_{i,j},\quad i,j\in Z
\ee
and act in the fermionic Fock space where the (right) vacuum vector $|0\rangle$
is annulated by each of $\psi^\dag_i,\psi_{-i-1},\,i\ge 0 $. In the Dirac sea picture we
consider that in the Dirac sea all states below sea level are occupied by the fermions 
$\psi_{-1},\psi_{-2},\dots.$

We take
\be
\deg \psi_n = n,\quad \deg \psi^\dag_n = -n
\ee

The basis elements $E_{i,j},\,i,j\in Z$ of the $\widehat{gl}_\infty$ algebra of the (generalized Jacobian) infinite matrices with entries $(E_{i,j})_{i',j'}=\delta_{i,i'}\delta_{j,j'}$ are realized by the normaly ordered bilinear Fermi operators $:\psi_i\psi^\dag_j:=\psi_i\psi^\dag_j-\langle 0|\psi_i\psi^\dag_j|0\rangle$. The central extention of the algebra of the infinite matrices ${gl}_\infty$ is defined by the cocycle 
\be\label{central}
\omega(E_{i,j},E_{j',i'})=
\begin{cases}
\delta_{i,i'}\delta_{j,j'},\quad {\rm if}\,\, i < 0,j \ge 0  
\\
0,\quad {\rm otherwise}
\end{cases}
\ee
and the charge $c=1$ 
\[
[a,b] \to [a,b]+c\omega(a,b),\quad a,b\in gl_\infty
\]
and in the fermionic realization
it is provided by the symbol of the normal ordering. For details see, for instance \cite{JM}.

A special role in the Kyoto school approach play the modes of fermionic current algebra
\be\label{J}
J_n=\sum_{i\in Z} :\psi_i\psi^\dag_{i+n}:,\quad n\in Z
\ee
As one can see $J_n|0\rangle =0$ and $\langle 0|J_{-n}=0$ for $n>0$.

The current operator $J_n$ is related to the matrix
$\sum_{i\in Z} E_{i,i+n}$ which can be vied as the diagonal matrix where nonvanisheng
entries are located $n$ positions above the main diagonal.

From the commutation relations in $\widehat{gl}_\infty$ one can see that these modes satisfy relations of the Heisenberg algebra
\be\label{Hei}
[J_n,J_{n'}]=n\delta_{n,n'},\quad n,n'\in Z
\ee
The right hand side is the result of the central extention, see (\ref{central}).
\br\label{about_central_extention}
Perhaps, the most evident way to understand the number $n$ in right hand side (\ref{Hei}) is to consider
the action of $J_{-n},\,n>0$ on the right vacuum vector looking at (\ref{J}),
namely at $J_{-n}|0\rangle$. This action pick up and move $n$ fermions $\psi_{-1},\dots,\psi_{-n}$ from the Dirac sea respectively to the positions $\psi_{n-1},\dots,\psi_0$ above the Dirac sea level
by the action of the terms $\psi_{n-1}\psi^\dag_{-1},\dots,\psi_{0}\psi^\dag_{-n}$ in (\ref{J}) (other terms
elliminate the vacuum).
While the consequent action of $J_n$ place these $n$ fermions back at their places. Thus,
$\langle 0|J_nJ_{-n}|0\rangle =n$ while $\langle 0|J_{-n}J_{n}|0\rangle =0$.
\er

Two commuting parts of the Heisenberg algebra consisting of $\{J_n,\,n>0\}$ and  $\{J_n,\,n<0\}$ 
are used to form evolutionary operators
\be
\Gamma_{\pm}(\pb_\pm)=\exp\sum_{n>0} \frac1n p_{\pm n} J_{\pm n},
\ee
where $\pb_\pm=(p_{\pm 1},p_{\pm 2},\dots)$ are sets of parameters called KP higher times,
which are used in the fermionic construction of the KP and 2KP tau functions:
\be\label{tauKP}
\tau^{\rm KP}_g(\pb_+)=\langle 0|\Gamma_+(\pb_+) g|0\rangle
\ee
\be\label{tau2KP}
\tau^{\rm 2KP}_g(\pb_+,\pb_-)=\langle 0|\Gamma_+(\pb_+) g\Gamma_-(\pb_-)|0\rangle
\ee
Here $g$ can be presented as an exponential of the elements of the $\widehat{gl}_\infty$ and 
in this sense (under some restrictions on the choice of $\widehat{gl}_\infty$) can be viewed as an element of the group of infinite matrices with the central extention. The choice of the fermionic operator $g$ defines
the choice of the tau function. Let us note that $u(\pb_+)=2\partial_{p_1}\log\tau^{\rm KP}(\pb_+)$ solves 
the equations of the famous KP hierarchy.

Let us note that $g\Gamma_-(\pb_-)$ in (\ref{tau2KP}) can be viewed as an example of $g$ in (\ref{tauKP})
which depends on the set of parameters $p_{-1},p_{-2},\dots$.

Let us recall that the tau function is an arbitrary chosen function of any two higher times, say, $p_1,p_2$ and the dependence
on other higher times are given by the evolution equation which are equations of the famous KP hierarchy of integrable equation.
We send the readers to the works of the Kyoto school for details.

In this paper we address the question what do we get if we replace $\Gamma_+(\pb_+)$
or $\Gamma_-(\pb_-)$, or the both evolutionary operators by the exponentials of elements of
different
abelian subalgebras of $\widehat{gl}_\infty$. In spite of the fact that it is quite natural
question which was mentioned in \cite{Orlov1988} it was not studied except of the cases considered 
in \cite{OS} and in \cite{O-rational-soliton}.

\section{Irreducible hierarchies of commuting operators}

\subsection{Generelized currents $J_{-n},\,n>0$}

\paragraph{Irreducible functions on $Z$}

Let us consider a function $r$ on the one-dimensional lattice $Z$. However,
we prefer to consider $r$ as a function of one variable on the complex plane $C$, which can be restricted on $Z$.

A given such a function $r$,  let us define the following analogue
of it's $k$-th root $\rho$ (one can call it quantum root of order $k$)
\be\label{quantum_root}
r(i) =\rho(i)\rho(i-1)\cdots \rho(i-k+1),\quad k=1,2,3\dots,
\ee
if it exists. As we see it does not exist in case $r$ has either an isolated zero,
or $k'$ consequent zeries, where $k'<k$. As we also see, any, say $s$ consequent 
zeroes of $\rho$ results in at least $k+s$ zeroes of $r$.

For a given $k$ we call $r$ {\it $k$-reducible} if there exists a solution to equation (\ref{quantum_root}). Otherwise, we call $r$ {\it $k$-irreducible}. Is $r$ $k$-irrreducible,
or it is not the case depends only on the location of zeroes of $r$ and the number $k$.

{\bf Example 1.} 
\be
r(x)=x^n,\quad n=1,2,3,\dots
\ee
is $k$-irreducible. 
While $r(x-a)$ for non-integer $a$ is $k$-reducible with
$\rho(x)=\frac{\Gamma(\frac{x-a-1}{k})}{\Gamma(\frac{x-a}{k})}$ be the solution to (\ref{quantum_root}).

{\bf Example 2.}
\be
r(x)= \prod_i^n(x-a_i)^{m_i}
\ee
is $n$-reducible for $a_i=i+\alpha,\,i=1,\dots,n$, $\alpha\in C$ with $\rho(x)=x-a_1$. It is also
reducible in case each $a_i$ is non-integer.
And it is $n$-irreducible, for instance, 
in case $a_i=1,\,i=1,\dots,n-1$ and $a_n\neq 0,n$.

\paragraph{Irreducible graded hierarchies of commuting operators}

Here we will consider $\widehat{gl}_\infty$ operators 
parametrized by an integer $p\neq 0$ and by a function $r$ on $Z$
as follows
\be\label{A}
A_{np}(r)=\sum_{i\in Z} r(i)r(i-p)\cdots r\left(i-p(n+1)\right):\psi_i\psi^\dag_{i-pn}:,\quad n,p =1,2,\dots 
\ee
These operators generalize currents $J_{-pn}$
\be\label{J_pn}
J_{-np}(r)=\sum_{i\in Z} :\psi_i\psi^\dag_{i-pn}:,\quad n,p=1,2,\dots
\ee

  If we take $\deg \psi_i=i,\,\deg \psi^\dag_i=-i$, 
and $\deg r(i)=0$ then  $\deg J_{-np}=\deg A_{np}(r)=-np$. The set $\{A_{np}(r),n>0\}$ generalizes the 
set $\{J_{-pn},n>0 \}$.
The difference between $A_{p}(r)$ and $J_{-p}$ is the prefactor $r(i)$.
The set of commuting operators $\{ J_{-np},n>0 \}$ is the subset of the larger sets of commuting operators $\{ J_{-n},n>0 \}$. The similar statement can be either valid, or not valid depending
of the location of zeroes of the function $r$ and the integer $p$.
We call the hierarchy of operators either reducible or irreducible depending on is there exists 
an integer $k$ such that $r$ which enter (\ref{A}) is either $k$-reducible or $k$-irreducible.
For a given $p$ and $r$ one can get a set $K$ of different numbers $k$ such that $r$ is $k$-reducible.
The reducible hierarchy $\{A_{pn}(r),\,n=1,2,3,\dots\}$ is a subhierarhy of the hierarchy 
$\{A_{qn}(\rho),\,n=1,2,3,\dots\}$ where $q={\rm min}k,\,k\in K$ and where $\rho$ is the 
the quantum root of $r$ of order $q$.

In what follows we will consider only irreducible hierarchies $\{A_{pn}(r),\,n=1,2,3,\dots\}$,
namely hierarchies which are not subhierarchies of another commuting sets 
$\{A_{qn}(\rho),\,n=1,2,3,\dots\}$, where $q<p$ (this unequality is important). 

\br
In the next section
we will show that for $p>1$ the hierarchy of commuting operators $\{A_{pn}(r),\,n=1,2,3,\dots\}$
is a subhierarchy of different commuting operators $\{A_{pn}(r'),\,n=1,2,3,\dots\}$, where $r'$ is a certain $p$-parametric deformation of $r$. 
\er

{\bf Example 3.} An example of (\ref{A}) with $n=1$ is the element of the Virasoro algebra $L_{p}$:
\be
A_p{r}=L_{-p} + NJ_{-p} =\sum_{i\in Z} (i+N)\psi_i\psi^\dag_{i-p}
\ee
where $r(i)=i+N$. The subcase $p=-1$ is of use in the context of the two-matrix model \cite{HO2003}
where $N$ plays the role of the matrix size. 

{\bf Example 4}. The other example is the following element of $W_{1+\infty}$ algebra
\be
A_p(r)=\sum_{i\in Z} (i+N)^n\psi_i\psi^\dag_{i-p},
\ee
which also $p$-irreducible. The case $r(i)=(N+i)^n$ and $p=1$ is used for the Ginibre ensemble
of the $n$ complex matrices. 

\br

This is the case where one can send $r$ to a canonical form 
\be
r(i)\to\theta(i)
\ee
where $\theta$ is either $1$ or $0$ (one can call $\theta$ characteristic function).

Namely there exists a diagonal matrix $T$, such that
\be
T^{-1} A_{np}(r) T=A_{np}(\theta)
\ee
Here
\be
T=\exp \left(\sum_{i < 0} T_i \psi_i\psi^\dag_i - \sum_{i \ge 0} T_i \psi^\dag_i \psi_i \right)
\ee
and 
\be
r(i)=e^{T_{i-1}-T_i}
\ee

\er

\br
From the point of view of the algebra $\hat{gl}_\infty$ realized as the algebra of infinite matrices,
the operator $A_{1;p}(r)$ is related to the infinite diagonal matrix ${\cal{A}}_{1;p}(r)$ whose diagonal is placed on the $p$ steps above (below) the main diagonal in case $p>0$ (in case $p<0$) and 
$\left({\cal {A}}_{1;p}(r)\right)_{i,j}=r(i)\delta_{i,i+p}$.
Then the matrix related to $A_{np}(r)$ is the $n$-th power of ${\cal {A}}_{1;p}(r)$.
Then if we have an isolated zero on the diagonal caused by $r(i-1)\neq 0,\, r(i)=0,\,r(i+1)\neq 0$
then the $n$-th power yields $n$ consequent zeroes.
\er

\paragraph{Colored partitions and $p$-component fermions.}

The structure of an operator
\be
\sum_{i} r(i) \psi_i\psi^\dag_{i-p}
\ee
prompts to split the set $Z$ of indices $i$ into the subsets classes modulo $p$.
So, let us introduce multicomponent fermions in a way it was done in \cite{JM}:
\be
\psi^{(c)}_i=\psi_{ip+c},\quad \psi^{\dag(c)}_i=\psi^\dag_{ip+c},\quad c=0,\dots,p-1
\ee
We get 
\be
[\psi^{(c)}_i,\psi^{(c')}_j]_+=0=[\psi^{\dag(c)}_i,\psi^{\dag(c')}_j],\quad
[\psi^{(c)}_i,\psi^{\dag(c')}_j]=\delta_{c,c'}\delta_{i,j},\quad i,j\in Z
\ee
Then, each basis Fock vector $|\lambda\rangle$ can be decomposed as the direct product
of states $|\lambda^{(c)}\rangle$, where each $\lambda^{(c)}$ is defined as follows:

We introduce coordinated of Fermi particles which yield  the basis vector $|\lambda\rangle$
by the relation 
\be\label{x-lambda}
x_i:=\lambda_i-i.
\ee
Then each coordinate $x_i$ has a color $c$ according to the rule 
\be
x_i=px^{(c)}_{j(i)}+c,\quad i=1,\dots,\ell(\lambda),\quad c=0,\dots,p-1,
\ee
where $\ell(\lambda)$ denotes the length of the partition $\lambda$, or the same, the number
of non-zero parts of $\lambda$. The values of $j$ are to be defined, and $j(i)>j(i-1)$. 
Then we have the set of $p$ partitions $\{\lambda^{(c)},\,c=0,\dots,p-1\}$ defined by
\be\label{x-lambda-c}
x^{(c)}_i:=\lambda^{(c)}_i-i
\ee

Example: let $p=3$ and $\lambda=(5,5,1)$. Then 
$\ell(\lambda)=3$ and according to (\ref{x-lambda}) we have three coordinates
$$
x_1=5-1=4,\,x_2=5-2=3,\,x_3=1-3=-2.
$$ 
We have $c=0,1,2$, and we can define the colors of the coordinates and there subscript label:
$$
x_1=3 x^{(1)}_1+1,\,
x_2=3 x^{(0)}_1+0,\,
x_3=3 x^{(1)}_2+1.
$$
As we see, 
the coordinate $x_2$ has a color $c=0$ and $x^{(0)}_1=1$.
The coordinates $x_1$ and $x_3$ have the color $c=1$, and $x^{(1)}_1
=1$, $x^{(1)}_2=-1$.
And there no coordinates with the color $c=2$.
Thus, according to (\ref{x-lambda-c}), we get $\lambda^{(0)}=(0)$, because $\lambda^{(0)}_1=x^{(0)}_1+1=3$. Next,
$\lambda^{(1)}$ with two parts, $\lambda^{(1)}=(3,3)$, because $(\lambda^{(1)}_1=2+1$ and $ \lambda^{(1)}_2)=1+2$, see (\ref{x-lambda-c}). At last, the partition $\lambda^{(2)}$ is empty.

\paragraph{Wider commutative hierachy.}
Now one can observe that we have more opportunities to get abelian subalgebras.
We have
\be
A_{np}(r)=\sum_{c=0}^{p-1}A^{(c)}_{n;1}(r^{(c)}),
\ee
where
\be
A^{(c)}_{n;1}(r^{(c)})=\sum_{i\in Z} r^{(c)}(i)r^{(c)}(i-1)\cdots r^{(c)}(i-n+1)\psi^{(c)}_{i}\psi^{\dag(c)}_{i-n}
\ee
and $r^{(c)}(i),\,i\in Z$ is the following set of functions
\be\label{r-r-c}
 r^{(c)}(i)=r(pi+c)
\ee
Therefore one can write
\be\label{Gamma-factors}
\Gamma_{p;r}(\pb):=\exp\sum_{n>0} \frac 1n p_n A_{np}(r)=
\prod_{c=0}^{p-1}\Gamma^{(c)}_{1;r^{(c)}}(\pb)
\ee

\paragraph{Matrix elements of the abelian multiparametric group}

We are interested in
$$
\langle \lambda| \Gamma_{p;r}(\pb)|\mu\rangle
$$
Using (\ref{Gamma-factors}) and the results of \cite{OS} we can write
\be\label{matrix-elements}
\langle \lambda|\Gamma_{p;r}(\pb)|\mu\rangle = r_{\lambda/\mu}s_{\lambda/\mu}(\tilde\pb)
\ee
\be
=\prod_{c=1}^p r^{(c)}_{\lambda^{(c)}/\mu^{(c)}}s_{\lambda^{(c)}/\mu^{(c)}}(\pb)
\ee

\section{Generalized currents $J_n,\,n>0$}
In the similar way, for a given $k$ and a given function $\tilde{r}$ we consider
\be\label{quantum_root'}
\tilde{r}(i) =\tilde{\rho}(i)\tilde{\rho}(i-1)\cdots \tilde{\rho}(i-k+1),\quad k=1,2,3\dots,
\ee
and call $\tilde{r}$ $k$-irreducible if there no $\tilde{\rho}$ which solves (\ref{quantum_root'}).
 
For a given $p$-irreducible $\tilde{r}$ we present the hierarchy 
of commuting operators 
\be\label{A-tilde}
\tilde{A}_{np}(\tilde{r})=\sum_{i\in Z} \tilde{r}(i)\tilde{r}(i+p)\cdots \tilde{r}\left(i+p(n-1)\right):\psi_i\psi^\dag_{i+pn}:,\quad n\in Z
\ee

In the same way we obtain
\be
\tilde{\Gamma}_{p;\tilde{r}}(\pb)=\exp\sum_{n>0} \frac 1n \tilde{p}_n \tilde{A}_{np}(\tilde{r})
\ee

For a given $\tilde{p}$ and $\tilde{\pb}=(\tilde{p}_1,\tilde{p}_2,\dots)$ we define
\be
\label{p-tilde-p}
\tilde\pb=(\underbrace{0,\dots,0}_{p-1},\tilde{p}_1,\underbrace{0,\dots,0}_{\tilde{p}-1},\tilde{p}_2,\dots)
\ee
and
\be
\tilde{r}_{\lambda/\mu}=\prod_{(i,j)\in \lambda/\mu} \tilde{r}(pj-pi)
\ee
\bl
We have
\be
\langle \mu|\tilde{\Gamma}_{\tilde{p};\tilde{r}}(\tilde{\pb})|\lambda\rangle = \tilde{r}_{\lambda/\mu}s_{\lambda/\mu}(\tilde\pb)
\ee
\be
=\prod_{c=1}^p \tilde{r}^{(c)}_{\lambda^{(c)}/\mu^{(c)}}s_{\lambda^{(c)}/\mu^{(c)}}(\tilde{\pb})
\ee
where $\tilde{r}$ and $\tilde{r}^{(c)}$ are related by (\ref{r-r-c}) and $\pb$ and $\tilde\pb$ are related by (\ref{p-tilde-p}).
\el

\section{Discussion}

This note is a more complete answer to a remark in the article \cite{Orlov1988} on commuting flows built on additional symmetries, and also develops \cite{OS}, where the exponent of the Abelian subalgebra is $W_{1+\infty}$ (namely, the one associated with the subalgebra $\{\partial_z^n\,,n>0$)
was applied to obtain a series of perturbations for a two-matrix model (see Section \ref{MatrMod} below) and for the one-matrix model \cite{HO2003}. In \cite{OS}, the role of the zeros of the $r$ function is also indicated.

It would be interesting to explicitly construct differential equations in which the independent variables are the group parameters of abelian symmetries, and analyse these equations.
In particular to get analogues of Leznov-Savel'ev open Toda lattices and
 to obtain analogues of open Todov chains of Leznov-Saveliev \cite{LeznovSaveliev} and also semi-open (``forced'') Toda chains \cite{GMMMO}. Let us note that they will posses symmetries related
 to multicomponent KP flows.

It is interesting to relate this study to the interesting works \cite{MM}, \cite{MMMPWZ}, \cite{MM-1}, \cite{MMP} (and also \cite{DMP}.
In certain sense it can be viewed as creation-annihilation point of view at the coherent states
formed by abelian subalgebras of $W_{1+\infty}$ algebra.)

The similar example with the BKP hierarchy was considered in \cite{MMNO} and will be considered
in more details in the next article.

\section{Acknowledgements}

The present work is an output of a research project implemented as part of the Basic Research Program at the National Research University Higher School of Economics (HSE University).

\appendix

\section{KP and $W_{1+\infty}$}

In this section we recall and re-write certain known facts about KP symmetries. We consider abelian
subgroups and their action on vacuum. Examples contain three well-known matrix models.

\subsection{Notations}

We denote the charged free fermions $\psi_i$ and $\psi^\dag_i$
\be
[\psi_i,\psi^\dag_j]_+=\delta_{ij},\quad [\psi_i,\psi_j]_+=[\psi^\dag_i,\psi^\dag_j]_+=0,\quad i,j\in\mathbb{Z} .
\ee
For a given Dirac sea level $n$ we have
\be\label{charged_canon}
\psi_i|N\rangle = \psi^\dag_{-i-1}|N\rangle  =0 = \langle N|\psi^\dag_i =\langle N|\psi_{-1-j},\quad i<N .
\ee
The charged fermionic fields
\be
\psi(z) =\sum_{j\in\mathbb{Z}} \psi_{j} z^{j-\frac12} ,\quad 
\psi^\dag(z) =\sum_{j\in\mathbb{Z}} \psi^\dag_{-j} z^{j-\frac12} .
\ee
In case we were interested to change the variable $z$, both fermionic fields transform 
as semi-forms $(dz)^{\frac 12}$, see \cite{GrinOrl}.

Let $\kappa$ and $\pb=(p_1,p_2,\dots)$ be a set of parameters.
The vertex operators
\bea\label{X=exp}
X(z)&=&e^{\sum_{i>0} \frac 1i z^i p_i } e^\kappa z^{\partial \kappa -\frac12} e^{-\sum_{i>0}\ z^{-i}\partial_{p_i}}
=  \sum_{i\in\mathbb{Z}} z^{i-\frac12}X_i    
,\\ 
X^\dag(z)&=&e^{-\sum_{i>0} \frac 1i z^i p_i } e^{-\kappa} z^{\frac12 -\partial \kappa}
e^{\sum_{i>0}  z^{-i}\partial_{p_i}} =\sum_{i\in\mathbb{Z}} z^{i-\frac12}X^\dag_{-i} 
\eea
(where $z^{\partial_\kappa}$ is the shift operator: 
$z^{\partial_\kappa} e^{\kappa}=z e^\kappa e^{\partial_\kappa}$)
act in the bosonic Fock space, which consists on polynomials in the variables $p_1,p_2,\dots$ times $e^{\kappa N}$:
\be
{\rm Pol}(p_1,p_2,\dots)e^{\kappa N},
\ee
where $\kappa$ is a formal parameter and the set $\{t_i:=\frac 1i p_i,\,i>0\}$ is called the set of 
KP higher times. The integer variable $N$ is the discrete time variable of the so-called modified KP and also the lattice variable of the relativistic Toda lattice \cite{Mikhailov},\cite{UT}, which can be viewed as a certain KP symmetry.
The anti-commutation relations of $X(z)$ and of $X^\dag(z)$ coinside with the anti-commutation relations 
(\ref{charged_canon}), where $\psi(z)$ is replaced by $X(z)$ and $\psi^\dag(z)$ is replaced by $X^\dag(z)$.

Formula (\ref{X=exp}) sometimes is written as
\be
X(z)=\vdots e^{\varphi^{\rm b}(z)} \vdots,\quad 
X^\dag(z)=\vdots e^{-\varphi^{\rm b}(z)} \vdots\,,
\ee
where 
\be\label{varphi}
\varphi^{\rm b}(z)=
\kappa +(\partial_\kappa -\tfrac12)\log z + \sum_{i>0}  \frac 1i z^i p_i - \sum_{i>0}  z^{-i}\partial_{p_i} \,,
\ee
and
where $\vdots A \vdots$ means that each shift-operator $e^{\pm \partial_{p_i}},\,i>0$ 
is moved to the right of the factor $e^{\mp \frac 1i z^i p_i }$  and the 'zero mode' shift operator 
$z^{\pm \partial_\kappa}$ is moved to the right of the factor $e^{\pm\kappa}$.

\paragraph{Currents.}
Consider
\be\label{curr-generator}
:\psi(z)\psi^\dag(z): = \sum_{m\in\mathbb{Z}} z^{-m-1} J^{\rm f}_{m} 
\ee
Symbol $:A:$ denotes the fermionic normal ordering, which, for $A$ a bilinear in the fermions $A$, can be equated
to $A-\langle 0|A|0\rangle$. 

Operators
\be\label{curr}
J^{\rm f}_m=\sum_{i\in\mathbb{Z}} :\psi_i\psi_{i+m}: =\res_z z^{m} :\psi(z)\psi^\dag(z): dz
\ee
are called fermionic currents.

As one can see 
\be
[J^{\rm f}_k , J^{\rm f}_m ]= k \delta_{m+k,0}
\ee
and
\be
J^{\rm f}_0|N\rangle =|N\rangle N,\quad z^{J^{\rm f}_0}|N\rangle =|N\rangle z^{N} .
\ee
One can introduce the operator $J^{\rm f+}_0$ by   
\be
e^{J^{\rm f+}_0}\psi_i=\psi_{i+1}e^{J^{\rm f +}_0},\quad
e^{J^{\rm f +}_0}\psi^\dag_i=\psi^\dag_{i+1}e^{J^{\rm f +}_0}
\ee
and by
\be
e^{kJ^{\rm f +}_0}|N\rangle = |N+k\rangle\,,
\ee
which result in 
\be
[J^{\rm f +}_0,J^{\rm f}_m]=-\delta_{m,0}.
\ee

 The Fermi fields can be written as
\bea\label{psi_via_J}
\psi(z)&=&e^{\sum_{m>0} \frac 1m z^m J^{\rm f}_{-m}} e^{J^{\rm f +}_0} z^{J^{\rm f}_0-\frac12}e^{-\sum_{m>0} 
\frac 1m z^{-m} J^{\rm f}_{m}} \left(dz\right)^\frac 12,\\
\psi^\dag(z)&=&e^{-\sum_{m>0} \frac 1m z^m J^{\rm f}_{-m}} e^{-J^{\rm f +}_0}z^{-J^{\rm f}_0+\frac12}
e^{\sum_{m>0} \frac 1m z^{-m} J^{\rm f}_{m}}\left(dz\right)^\frac 12
\eea
or, it can be written as
\be
\psi(z)=\vdots e^{\varphi^{\rm f}(z)}\vdots ,\quad \psi(z)=\vdots e^{-\varphi^{\rm f}(z)}\vdots\,,
\ee
where
\be
\varphi^{\rm f}(z)= 
J^{\rm f +}_0+J^{\rm f}_0\log z + \sum_{m>0} \frac 1m z^m J^{\rm f}_{-m} - \sum_{m>0} \frac 1m z^{-m} J^{\rm f}_m\,,
\ee
and where 
 $\vdots A \vdots$ means that the currents $J^{\rm f}_i,\,i>0$ 
are moved to the right of the currents  $J^{\rm f}_i,\,i< 0$  
while 'zero mode' $J^{\rm f +}_0$ is moved to the left of $J^{\rm f}_0$.

The formula of this type was first discovered in the work of Pogrebkov and Sushko \cite{PogrebkovSushko}
\footnote{The preprint of this work was not published in journal version for long, because referees decided that
it is too unusual and can be wrong.}

Bosonic currents are defined as
\be
J^{\rm b}_m=\begin{cases}
           m\partial_{p_m},\quad m>0\\
           p_0,\quad m=0\\
            p_{-m},\quad m<0
          \end{cases}\,.
\ee
\br
As one can verify this definition is equivalent to
\be
J^{\rm b}_m
=\lim_{y\to 0}\res_z z^{m}\left(\vdots X(z(1+\tfrac y2))X^\dag(z(1-\tfrac y2)) \vdots -1\right)\frac{dz}{y}
=\res_z z^{m} \vdots \left(D\cdot X(z)\right) X^\dag(z) \vdots \frac{dz}{z}\,,
\ee
where $\vdots A \vdots$ (which means that the shift-operators $e^{\pm \partial_{p_i}},\,i>0$ of the both vertex operators
are moved to the right and $e^{\pm\kappa}$ are moved to the right),  according to the Campbell-Hausedorff formula,
is results in the appearence of the factor $\left(1-\tfrac{1-\frac y2 }{1+\frac y2 }\right)^{-1}=\frac 1y +O(1)$.
\er

\paragraph{ Fermion-boson correspondence}:
\bea
\psi(z) &\longleftrightarrow & X(z) \label{FB1}\\
\psi^\dag(z) &\longleftrightarrow & X^\dag(z)\label{FB2}\\
|N\rangle &\longleftrightarrow & e^{\kappa N}\label{FB6}  
\eea
In particular, we have
\bea
R &\longleftrightarrow & e^\kappa\label{FB4}  \\
J^{\rm f}_m &\longleftrightarrow & J^{\rm b}_m\label{FB3}\\
\varphi^{\rm f}(z)&\longleftrightarrow & \varphi^{\rm b}(z)\\
J^{\rm f}_{-\lambda_1}\cdots J^{\rm f}_{-\lambda_k}|N\rangle &\longleftrightarrow & 
 \label{FB7} e^{\kappa N} p_{\lambda_1}\cdots p_{\lambda_k}
\eea

We shall omit the superscripts ${\rm f}$ and ${\rm b}$ and hope that it does not produce a mess.

\subsection{$W_{1+\infty}$ algebra and it's abelian subalgebras \label{W}}

Let us use few notions from the textbook \cite{Mac}.
Parition is a non-increasing set of nonnegative integers, say $\lambda=(\lambda_1,\dots,\lambda_l)$,
$\lambda_i\ge\cdots\ge\lambda_l\ge 0$. The sum $\sum_i \lambda_i =:|\lambda|$ is called the weight of $\lambda$.
The number of the nonvanishing parts of $\lambda$ is called the length of $\lambda$ and denoted by $\ell(\lambda)$.
The numbers $z_\lambda$ and $z'_\lambda$ are equal respectively to $\prod_{i} m_i!i^{m_i}$ and to
$\prod_{i} m_i!$   where $m_i$ is the number of times the integer $i$
occurs in $\lambda$. For $\lambda=0$ we put $z_0=z'_0=1$.  (The number $z'_\lambda$ will be used in (\ref{TaylorXX^dag}) below). We denote the set of all partitions (including zero partition) by $\Pa$.

There is the well-known relation
\be\label{e}
e^{\sum_{m>0} \frac 1m p_m\tilde{p}_m}=\sum_{\lambda\in\Pa} \frac{1}{z_\lambda} \pb_\lambda\tilde{\pb}_\lambda\,,
\ee
 where $\pb=(p_1,p_2,\dots)$ and $\tilde{\pb}=(\tilde{p}_1,\tilde{p}_2.\dots)$ are two (infinite) sets of parameters and
where $\pb_\lambda=\prod_{i=1}^{\ell(\lambda)}p_{\lambda_i},\,\,\tilde{\pb}_\lambda=\prod_{i=1}^{\ell(\lambda)}\tilde{p}_{\lambda_i}$.
It is assumed that $\pb_0=\tilde{\pb}_0=z_0=1$.

By this relation and by (\ref{psi_via_J}) one gets:
\be\label{psipsidag}
\psi(x)\psi^\dag(y) =\frac{1}{1-yx^{-1}}\frac 1x \left(\frac{x}{y}\right)^{J_0-\frac12}
\sum_{\lambda,\mu\in\Pa} \frac{1}{z'_\lambda z'_\mu}x^{-|\mu|} y^{-|\mu|} (x/y)'_\lambda (x/y)'_\mu
\Jb^{\rm f}_\lambda \Jb^{\rm f}_{-\mu}\,,
\ee
where 
\begin{itemize}
  \item $\Jb^{\rm f}_\lambda := \prod_{i=1}^{\ell(\lambda)} J^{\rm f}_{\lambda_i}$ and $\Jb^{\rm f}_0:=1$ 
  (don't miss with $J_0$)
  \item $\Jb^{\rm f}_{-\mu} := \prod_{i=1}^{\ell(\mu)} J^{\rm f}_{-\mu_i}$ and $\Jb^{\rm f}_{-0}:=1$
  \item $(x/y)_\lambda:=\prod_{i=1}^{\ell(\lambda)}(x^{\lambda_i}-y^{\lambda_i}) 
  \prod_{i=1}^{\ell(\lambda)}\frac{1}{\lambda_i} $ in case $\lambda\neq 0$ (the last product changes $z_\lambda$ 
  in (\ref{e}) to $z'_\lambda$ in (\ref{psipsidag})),
  and   $(x/y)_0:=1$
  \item $(x/y)_\mu:=\prod_{i=1}^{\ell(\mu)}(x^{\mu_i}-y^{\mu_i})\prod_{i=1}^{\ell(\mu)}\frac{1}{\mu_i} $ 
    in case $\mu\neq 0$   and 
  $(x/y)_0:=1$
\end{itemize}

\paragraph{I.}
Let us write $x=ze^{\frac y2}$, $x=ze^{-\frac y2}$ and assign $\deg z=1$.

Consider
\be\label{W^f}
:\psi(ze^{\frac y2})\psi^\dag(ze^{-\frac y2}):
 = \sum_{i,m\in\mathbb{Z}}z^{m-1}e^{(i-\frac m2) y}:\psi_i\psi^\dag_{j}: =
\ee
\be
 =
\sum_{m\in\mathbb{Z},\,n\ge 0}\frac{1}{n!}z^{m-1} y^n 
\sum_{i\in\mathbb{Z}}(i-\tfrac m2)^n:\psi_i\psi^\dag_{i-m}:
=:\sum_{m\in\mathbb{Z},\,n\ge 0} \frac{1}{n!} z^{m-1}y^n W^{\rm f}_{m,n}
\ee
In other words
\be
W^{\rm f}_{m,n}=\res_z : \left( z^{-\frac m2} D^n z^{-\frac m2} \cdot \psi(z)\right)\psi^\dag(z): dz
\ee
The set $\{ W^{\rm f}_{m,n},\,n\ge 0,\,m\in\mathbb{Z}\}$ is the set of the generators of the $W_{1+\infty}$
algebra.

The bosonic counterparts can be written down as follows. For practical calculations
in the bosonic Fock space one prefers to have normal ordered expressions.

There are two ways to present explicit formulas for it.
 In both cases,
it is convinient to take into account
that $:A:=A-\langle 0|A|0\rangle$ and write
 the bosonic counterpart of (\ref{W^f})
as follows:
\be\label{W-bosonic}
X(ze^{\frac y2})X^\dag(ze^{-\frac y2})-
\frac{1}{z e^{\frac y2}-z e^{-\frac y2}}
=:\sum_{m\in\mathbb{Z},\,n\ge 0} \frac{1}{n!} z^{m-1}y^n W^{\rm b}_{m,n}
\ee
where $\vdots A \vdots$ means that the shift-operators parts of the both vertex operators
are moved to the right which, according to the Campbell-Hausedorff formula, is equivalent to the 
appearence of the factor $\frac{e^{yp_0}}{1-e^{-y}}$.
Thus, this symbol $ \vdots \vdots $ means the bosonic ordering, in which all derivatives with respect to $ p_i $
variables are moved to the right of functions of $ p_i $ variables.

\paragraph{I} The first way to write down

$$
\vdots X(ze^{\frac y2})X^\dag(ze^{-\frac y2}) \vdots= e^{y\partial_\kappa}
e^{\sum_{k>0} \frac 1k z^k\left( e^{k\frac y2}- e^{-k\frac y2} \right) p_k }
e^{\sum_{k>0}  z^{-k}\left( e^{k\frac y2}- e^{-k\frac y2} \right) \partial_{p_k} } ,
$$

$$
X(ze^{\frac y2})X^\dag(ze^{-\frac y2}) -\frac{1}{ze^{\frac y2}-z e^{-\frac y2} } =
  \frac{1}{ze^{\frac y2}-z e^{-\frac y2} } \left(
\vdots X(ze^{\frac y2})X^\dag(ze^{-\frac y2}) \vdots  -1\right)
$$
\be\label{TaylorXX^dag}
=\frac{z^{-1}}{\sinh'_{(1)}(\tfrac y2)}\left(
 e^{y\partial_\kappa}
\sum_{\lambda,\mu\in\Pa} \frac{1}{z'_\lambda z'_\mu}
z^{|\lambda|-|\mu|}\sinh'_\lambda(\tfrac y2) \sinh'_\mu(\tfrac y2)\pb_\lambda\tilde{\partial}_\mu
-1\right)\,,
\ee
where
\begin{itemize}
  \item $\pb_\lambda := p_{\lambda_1}\cdots p_{\lambda_l}$, if $\lambda_l>0$. In case $\lambda=0$ we put $\pb_0:=1$
  \item $\tilde{\partial}_\mu=\left(\mu_1\partial_{p_{\mu_1}} \right)\cdots \left(\mu_k\partial_{p_{\mu_k}} \right) $, 
  where $\mu_k>0$. In case $\mu =0$ we put $\tilde{\partial}_0:=1$.
  \item $\sinh'_\lambda(\tfrac y2):=
 \sinh \left(\tfrac 12 \lambda_1 y \right)\cdots 
 \sinh \left(\tfrac 12 \lambda_1 y \right)\prod_{i=1}^{\ell(\lambda)}\frac{2}{\lambda_i}$ in case $\lambda_l>0$ and 
 $ \sinh'_\lambda(\tfrac y2):=1 $ in case $\lambda=0$
 \item $\sinh'_\mu(\tfrac y2):=
 \sinh \left(\tfrac 12 \mu_1 y \right)\cdots 
 \sinh \left(\tfrac 12 \mu_1 y \right)\prod_{i=1}^{\ell(\mu)}\frac{2}{\mu_i}$ and 
 $ \sinh'_\mu(\tfrac y2):=1 $ in case $\mu=0$
\end{itemize}
Let us replace $\partial_\kappa$ by it's eigenvalue $p_0$.

Let us introduce
\bea
&&e^{y p_0}\frac{\sinh'_\lambda(\tfrac y2)\sinh'_\mu(\tfrac y2)}{\sinh'_{(1)}(\tfrac y2)} y^{1-\ell(\lambda)-\ell(\mu)}
=: \sum_{i\ge 0} y^i f_i(\lambda,\mu,p_0) \\
&=& 1+y p_0 +
y^2\left(\tfrac 12 (p_0-\tfrac12)^2+\tfrac{1}{12}+
\tfrac{1}{24}\sum_{i=1}^{\ell(\lambda)}\lambda^3_i+\tfrac{1}{24}\sum_{i=1}^{\ell(\mu)}\mu^3_i  \right)+\cdots\,,
\eea
where $f_0(\lambda,\mu,N)\equiv 1$ and $f_1(\lambda,\mu,N)=N$.

In particular,
$$
\frac{y}{e^{\frac y2}-e^{-\frac y2}}=\sum_{i\ge 0} y^{i} f_i(0,0,0)
=1 +y^2 \frac{2}{2^2}\frac{1}{3!} + \cdots
$$
$$
\frac{y e^{y N}}{e^{\frac y2}-e^{-\frac y2}}=\sum_{i\ge 0} y^{i} f_i(0,0,N)
$$
Then
$$
\sum_{m\in\mathbb{Z},\,n\ge 0} \frac{1}{n!} z^{m} y^n W^{\rm b}_{m,n}=
$$
\be =
\sum_{m\in\mathbb{Z}} z^{m}
\sum_{\lambda,\mu\in\Pa \atop |\lambda|-|\mu|=m} \,
\frac{\pb_\lambda\tilde{\partial}_\mu}{z'_\lambda z'_\mu}\,
 \sum_{i\ge 0} y^{ \ell(\lambda)+\ell(\mu)-1 + i } f_i(\lambda,\mu,N)
 - \frac{1}{e^{\frac y2}-e^{-\frac y2}}\,,
\ee
which results in
\be\label{m=0}
 \frac{1}{n!} W^{\rm b}_{0,n}=\sum_{0\le i \le n+1}
 \sum_{ |\lambda|=|\mu| \atop \ell(\lambda)+\ell(\mu) + i=n+1} \,
 f_i(\lambda,\mu,N)\,
 \frac{\pb_\lambda\tilde{\partial}_\mu}{z'_\lambda z'_\mu}\,
   - f_{n+1}(0,0,0)  ,
\ee
where $\pb_0=\tilde{\partial}_0=z'_0= f_0=1$, and in
 $$
 \sum_{n\ge 0} \frac{1}{n!}  y^n W^{\rm b}_{m,n}
 =
 \sum_{\lambda,\mu\in\Pa \atop |\lambda|-|\mu|=m\neq 0} \,
 \frac{\pb_\lambda\tilde{\partial}_\mu}{z'_\lambda z'_\mu}\,
 \sum_{i\ge 0} y^{ \ell(\lambda)+\ell(\mu)-1 + i } f_i(\lambda,\mu,N),\quad m\neq 0.
 $$
or, the same
\be\label{m_not_0}
\frac{1}{n!}  W^{\rm b}_{m,n}
 =
\sum_{0\le i \le n}
\sum_{\lambda,\mu: |\lambda|-|\mu|=m\neq 0\atop \ell(\lambda)+\ell(\mu) + i=n+1} \,
f_i(\lambda,\mu,N)\,
 \frac{\pb_\lambda\tilde{\partial}_\mu}{z'_\lambda z'_\mu}\,,\quad m\neq 0.
 \label{TaylorXX^dag'}
\ee

As one can see for given $m,n$ we get a contribution of terms in the right hand side of (\ref{TaylorXX^dag'}) 
conditioned by 
\be\label{m}
|\lambda|-|\mu|=m
\ee
\be\label{n}
\ell(\lambda)+\ell(\mu)=  n+1 -i,\quad i=0,\dots,n+1
\ee

\br
For $m=0$, formula (\ref{m=0}) yields the generating (in the parameter $y$) function of the Hamiltonians of the quantum KdV
equation in free fermion point (see \cite{Dubr}, \cite{E} where it was written in another way).
\er

\br\label{i_terms}
The case $i=n+1$ is related to the $\lambda=\mu=0$ term (the 'free term')   in case $m = 0$ is   
$f_{n+1}(0,0,N)-f_{n+1}(0,0,0)$. Due to (\ref{m}) there is no free term in case $m\neq 0$
and in this case one studies $i=0,\dots,n$. In case $m\neq 0$ the terms where $i\ge 1$ exists only due to the 
prefactor which appears thanks to the bosonic ordering while in case $m=0$ there is the additional contribution
of the fermionic ordering which enters the free term for any given $n$.
That is why the summation ranges over $i$ in (\ref{m=0}) and in (\ref{m_not_0}) are different.

\er

 {\bf Examples}.

 (a) In (\ref{m=0}) $m=n=0$. Then we have two terms, $i=1$ and $i=0$ at most in the sum on the right hand side of (\ref{m=0}). 
 The case $i=0$ is impossible because the conditions $|\lambda|=|\mu|$ (which follows from $m=0$) 
 and the condition  $\ell(\lambda)+\ell(\mu) =1$ (which follows from $n=0$) 
 are inconsistent. Due to the $i=1$ term we get $W_{0,0}=f_1(N)-f_1(0)=N$.

 (b) In (\ref{m_not_0}) $n=0$ and $m\neq 0$. Then $i=0,1$. The case $i=1$ is impossible because condition (\ref{n}) 
 (where we get $\lambda=\mu=0$)
 is inconsistent with (\ref{m}) where $m\neq 0$.  Then, for $i=0$ we have either $\lambda=(m),\,\mu=0$ in case $m>0$,
 or $\lambda=0,\,\mu=(-m)$ in case $m<0$. 
 In both cases $z'_\lambda=z'_\mu=1$ and we get $W_{m,0}=J^{\rm b}_{-m}$.

 (c)  In (\ref{m=0}) take $n=1$ and $m=0$. In this case $i=0,1,2$.
 The terms with $i=2$ and $f_{i=2}(0,0,0)$ form the free term
 $f_2(0,0,N)-f_2(0,0,0)= \tfrac 12 (N - \tfrac12)^2$.
 The terms $i=1$ does not contribute because (\ref{m}) and (\ref{n}) are inconsistent for $i=1$.
 The terms $i=0$ gives the sum over $k\ge 1$ over partitions $\lambda=\mu=(k)$ and in this case
 $z'_\lambda=z'_\mu=1$.
  We obtain
 \be
 W_{0,1}= \tfrac 12 (N - \tfrac12)^2
 + \sum_{k>0} k p_k \partial_{p_k} =: L_0
 \ee

 (d) Take $n=1$ and $m=|\lambda|-|\mu|\neq 0$. In this case $i=0,1$; see Remark \ref{i_terms}.
 The contribution of $i=1$ terms results in either $\lambda=(m),\, \mu=0$ (for $m>0$), or in
 $\lambda=0,\, \mu=(-m)$ (for $m<0$), thus, it is equal to $f_{i=1}(\lambda,\mu,N)J^{\rm b}_{-m}$.
 The contribution of $i=0$ terms consists of two groups. The first group is related to  $\lambda=(k,m-k)$
 (if $k\ge m-k$) and to $\mu=0=$ in case $m>0$ and to $\mu=(k,m-k)$
 (if $k\ge m-k$) and to $\lambda=0$ in case $m<0$. The second group is related to 
 $\lambda=(k),\,\mu=(m-k)$. Taking into account $(z'_\lambda z'_\mu)^{-1}$ factor we finelly get 
 $$
 W_{m,1}=
 $$
 \be\label{W_virasoro}
 \begin{cases}
         N p_{m}+\sum_{k>m} p_k (k-m)\partial_{p_{k-m}}
 +\tfrac12\sum_{0< k<m }  p_kp_{m-k},\quad m > 0\\
 \tfrac 12 (N-\tfrac12)^2 + \sum_{k>0} k p_k \partial_{p_k},\quad m=0\\
 -m N\partial_{p_{-m}}+\sum_{k>0} p_k (k-m)\partial_{p_{k-m}}
 +\tfrac12\sum_{0< k < m} k(-m-k)\partial_{p_k}\partial_{p_{-m-k}} ,\quad m< 0
         \end{cases} 
 \ee
 where $\tfrac12$ aprior the second sums appear from the restriction $k\ge m-k$ in partitions $(k,m-k)$ 
 and separately from the factors $z'_\lambda$ and $z'_\mu$  for partitions $(k,k)$.
 
\br The combination
\be
L_{m}:=W_{-m,1}-N W_{-m,0}
\ee
coincides with the Virasoro algebra element $L_m^j$ presented in formula (3.7) of \cite{GrinOrl} where we put $j=\tfrac12$.

\er
 
 \paragraph{II} it is convinient to take into account
 that $:A:=A-\langle 0|A|0\rangle$ and write

$$
X(ze^{\frac y2})X^\dag(ze^{-\frac y2}) -\frac{z^{-1}}{1-e^{-y}} =
\frac{z^{-1}}{1-e^{-y}}\left(
\vdots X(ze^{\frac y2})X^\dag(ze^{-\frac y2}) \vdots e^{-\frac y2} -1\right)
$$
\be\label{W^b'}
=:\sum_{m\in\mathbb{Z},\,n\ge 0} \frac{1}{n!} z^{m-1}y^n W^{\rm b}_{m,n}\,,
\ee
where $\vdots A \vdots$ means that the shift-operators parts of the both vertex operators
are moved to the right which, according to the Campbell-Hausedorff formula, is equivalent to the 
appearence of the factor $\frac{z^{-1}e^{-\frac y2}}{1-e^{-y}}$.
Thus, this symbol $ \vdots \vdots $ means the bosonic ordering, in which
all derivatives with respect to $ p_i $
variables are moved to the right of functions of $ p_i $ variables.

Using
 $$
 \res_z z^{-m} \vdots \left(e^{\frac y2 D} X(z)\right)\left(e^{-\frac y2 D} X^\dag(z)\right)\vdots
 dz =\res_z\vdots \left(\left(e^{\frac y2 D} \cdot z^{-m}\cdot e^{\frac  y2 D}\right)\cdot X(z)\right)  X^\dag(z)\vdots
 dz 
 $$
 $$
 = \res_z z^{-m} \vdots  \left( e^{y (D-\frac m2)}\cdot X(z)\right)  X^\dag(z)\vdots
 dz 
 $$
we can rewrite (\ref{W^b})-(\ref{W^b'}) as follows:
$$
 W^{\rm b}_{m,n} = 
$$ 
\be 
 \res_{y=0} \frac{dy}{y}y^{n-1} \frac{y}{1-e^{-y}}
 \left\{e^{\frac{my}{2}}\sum_{k\ge 0} \frac{1}{k!}y^k
 \left[\res_z z^{-m }\vdots \left( D^k  \cdot X(z)\right)X^\dag(z)\vdots \frac{dz}{z}  \right]-\delta_{m,0}\right\}
\ee
\be\label{W^b}
=\res_{y=0} \frac{dy}{y}y^{n-1} \frac{y}{1-e^{-y}}
 \left\{e^{\frac{my}{2}}\sum_{k\ge 0} \frac{1}{k!}y^k
 \left[\res_z z^{-m }\vdots c_k(\varphi(z))\vdots \frac{dz}{z}  \right]-\delta_{m,0}\right\}
\ee
where $c_k(\varphi(z))=\left(D^k\cdot e^{\varphi(z)}\right)e^{-\varphi(z)}=
\left(zJ(z) \right)^k+\cdots+D^{k-1}\cdot zJ(z)$, $\varphi_z=J(z)$, $X(z)=\vdots e^{\varphi(z)}\vdots$; 
see (\ref{varphi}).
As we see the normally ordered bosonic expressions are more involved than normally ordered ferionic ones.

In what follows we omit the superscripts $^{\rm f}$ and $^{\rm b}$ and hope it does not produce a misunderstanding.

In particular the formula (\ref{W^b}) yields 
$$
W_{m,0}=\begin{cases}
         p_m,\quad  m>0\\
         N,\quad m=0\\
         -m\partial_{p_{-m}},\quad m<0
        \end{cases}
$$
and 
$$
W_{m,1}=2\begin{cases}
\sum_{k>0} \left( k \,p_{k-m}\partial_{p_{k}}+\frac12 k(m-k)\partial_{p_{k}}\partial_{p_{m-k}}\right),\quad m > 0\\
\sum_{k>0}  k \,p_k\partial_{p_k}+\frac 12 N^2,\quad m=0\\
     \sum_{k>0} \left( k \,p_{k-m}\partial_{p_k}+\frac 12 p_k p_{-k-m}\right)    ,\quad m<0
        \end{cases}
$$

\paragraph{III.} The third way is as it was done in \cite{Orlov1988} (in this work it was done
almost without examples). Consider
\be
:\psi(z+\epsilon)\psi^\dag(z):=\psi(z+\epsilon)\psi^\dag(z)-\frac{1}{\epsilon}=
:\sum_{m\in\mathbb{Z},\,n\ge 0} z^{m-1}\epsilon^n  \Omega^{\rm f}_{m,n}
\ee
One can write
\be
 \Omega^{\rm f}_{-m,n}=\frac{1}{n!}\res_z z^{m}:\frac{\partial^n\psi(z)}{\partial z^n}\psi^\dag(z): dz
\ee
\be
=\sum_{j\in\mathbb{Z}} (j-\tfrac 12)(j-\tfrac 32) \cdots (j-n-\tfrac 12)\psi_j \psi^\dag_{j+m}
\ee
In particular,
\be
\Omega_{-m,1}=
\sum_{j\in\mathbb{Z}}  j:\psi_j \psi^\dag_{j+m}:-\tfrac12 J_m
\ee
compare to
\be
W_{-m,1}=\sum_{i\in\mathbb{Z}}  i: \psi_i \psi^\dag_{i+m}:+\tfrac m2 J_m =:L^{j=\frac 12}_m
\ee
which is the element of the Virasoro algebra
\be
[L^j_m,L^j_n]=(n-m)L^j_{m+n}+(6j^2-6j+1)\frac{n^3-n}{12}\delta_{m+n,0}
\ee
$$
\sum_{i=1}^{n-1}( i^2 +  i)=\tfrac 13 (n^3-n)
$$
$$
m^2+\sum_{j=1}^{m-1} j^2 =  am^3 +(b+1)m^2+cm+d = a(m+1)^3+b(m+1)^2 + c(m+1)+d
$$
$$
a+b+c =0,\quad 0=3a+2b,\quad 1=3a
$$
$$
\sum_{j=1}^{m-1} j^2 = \tfrac 13 m^3 - \tfrac 12 m^2 +\tfrac 16 m = m(\tfrac 13 m^2 -\tfrac 12 m + \tfrac 16 )
$$
$$
\sum_{j=1}^{m-1} j = \frac{m(m-1)}{2}
$$

\be
:X(z+\epsilon)X^\dag(z):   =X(z+\epsilon)X^\dag(z)-\frac{1}{\epsilon}=
:\sum_{m\in\mathbb{Z},\,n\ge 0} z^m\epsilon^n  \Omega^{\rm b}_{m,n}
\ee

\be
=\lim_{\epsilon\to 0}\left( X(z+\epsilon)X^\dag(z)-\frac{1}{\epsilon} \right)
+\sum_{n\ge 1} \frac{1}{n!}\epsilon^{n} \frac{\partial^{n} X(z)}{\partial z^{n}}X^\dag(z)  
\ee
where the first term with $\lim$ contains vanishing $\epsilon^{-1}$ term and also the term $\epsilon^0$ 
(linear in currents, see below).

The right hand side is not convinient in the bosonic realization. It is convinient to write
\be
X(z+\epsilon)X^\dag(z)-\frac{1}{\epsilon} 
= \frac{1}{\epsilon} \left(\vdots X(z+\epsilon) X^\dag(z) \vdots -1\right)
\ee
\be
\sum_{n\ge 0} \frac{1}{n!}\epsilon^{n}\vdots \frac{\partial^{n+1} X(z)}{\partial z^{n+1}}X^\dag(z)  \vdots 
=:\sum_{m\in\mathbb{Z},\,n\ge 0} z^m\epsilon^n  \Omega^{\rm b}_{m,n}\,.
\ee
Or
\be
 \Omega_{m,n}=\frac{1}{n!}\res_{z=0} z^{m-1} \vdots \frac{\partial^{n+1} X(z)}{\partial z^{n+1}}X^\dag(z)  \vdots   dz
\ee
Then
\be
\Omega_{m,0}=J_{-m}\,,
\ee

\be
\Omega_{m,1}=\tfrac 12 \res_{z=0} z^{-m} \vdots  \left(\tfrac 12
\left(\sum_{k} z^{k-1}J_k\right)^2 \vdots+\sum_{k} (k-1)z^{k-2}J_k   \right) \frac{dz}{z}  . 
\ee

\paragraph{Graded elements which we  use.}
For our purposes,
the elements of the $W_{1+\infty}$ algebra in the bosonic Fock space will be \cite{KacRadul}  chosen as
\be\label{W_n_F}
W_{n}[F]=\res_z \left( z^n F(D) \cdot X(z)\right) X^\dag(z)\frac{dz}{z},\quad n\neq 0
\quad D:=z\frac{\partial}{\partial z} 
\ee
The case $n=0$ will be recalled separetely.
The pseudodifferential operator $F(D)$ acts on the formal series in the powers of $z$
according to the rule $F(D)\cdot z^k=F(k)z^k,\,k\in\mathbb{Z}$ where 
$F$ is a function on the lattice.  We consider it to be bounded
except the case $n=0$. Zeroes are admissible.

In the fermionic Fock space these are
\be
W^{\rm f}_{n}[F]=\res_z :\left( z^n F(D) \cdot \psi(z)\right) \psi^\dag(z): \frac{dz}{z}
\ee
\be
=\sum_{i\in\mathbb{Z}} F(i)\psi_i \psi^\dag_{i+n}
\ee
Here $:A:$ denotes $A-\langle 0|A|0\rangle$.

In case it does not produce a misunderstanding we shall omit the superscript ${\rm f}$ which says
that we deal with the fermionic version.

\begin{definition}
 For a given function $F$ on the lattice $\mathbb{Z}$, we
introduce the {\it characteristic function} $\mathfrak{F}[F]$ defined on $\mathbb{Z}$ which takes values $1$ and $0$ and whose zeroes
coinsides with the zeroes of $F$.
\end{definition}
Example: We take $\mathfrak{F}[x+n](x)=0$ in case $x=-n$, otherwise it is equal to 1.

We introduce the degree putting $\deg z=1$, then we get $\deg (W_n[F])=n$ independent of the choice of $F$.

One can verify the
\bl
For any choice of $F_1$ and $F_2$ we have
\be
\left[ W_{0}[F_1],W_{0}[F_2] \right]=
\left[ W^{\rm f}_{0}[F_1],W^{\rm f}_{0}[F_2] \right] =0
\ee
\el

\begin{proposition}
For a given $F$,
there exists $\mathbb{T}$ acting in the fermionic Fock space  such that
\be
 W^{\rm f}_n[F]= \mathbb{T} \,W^{\rm f}_n[\mathfrak{F}[F]] \,\mathbb{T}^{-1}
\ee
and  $\deg \mathbb{T}=0$.
\end{proposition}
(Thus, all $W^{\rm f}_n[F]$ with a given $\mathfrak{D}(F)$ belong to the same orbit.)

Proof. 
\be
W^{\rm f}_{n}[\mathfrak{F}]=\sum_{i} \mathfrak{F}[F](i)\psi_{i}\psi^\dag_{i+n}
\ee
Let
\be
\mathbb{T}=e^{ \sum_{i<0}T_i\psi^\dag_i\psi_i  -\sum_{i\ge 0} T_i\psi_i\psi^\dag_i  }=
:e^{W_0[T]}
\ee
where each $T_i$ is a finite number;
then 
\be\label{psi(T)}
\mathbb{T} \,\psi_i \,\mathbb{T}^{-1}=e^{-T_i}\psi_i,\quad 
\mathbb{T} \,\psi^\dag_i \,\mathbb{T}^{-1}=e^{T_i}\psi^\dag_i
\ee
Formaly, the set $\{T_i,\,i\in\mathbb{Z} \}$ can contain infinite numbers. Then one can get 0 in the right hand sides of relations (\ref{psi(T)}).

Thus,
$$
e^{T_{i+n} - T_{i}}\mathfrak{F}[F](i) = F(i)
$$
one can construct the set of $\{ T_i,\,i\in\mathbb{Z} \}$ with this property in a recurrent way.

\paragraph{Abelian subalgebras.}

For a given $n$ and $F$,
introduce the set
\be
J_m(n,F):=\res_z \left( \left(z^n F(D)\right)^m \cdot X(z)\right) X^\dag(z)\frac{dz}{z},\quad m=1,2,3,\dots
\ee
where $J_1(n,F)=W_n[F]$; see (\ref{W_n_F}).

\begin{proposition} For a given $F$ and a given $n\in\mathbb{Z}$, we have
\be
\left[ J_m(n,F),J_{m'}(n,F)\right]=0,\quad m,m'=1,2,3,\dots
\ee
\begin{remark}
In the BKP case below we have different situation: odd and even $n$ are rather different.
\end{remark}

\end{proposition}
Proof. We write
\be
J_m(n,F)=\sum_i F(i)F(i+n)F(i+2n)\cdots F(i+n(m-1)) \psi_i\psi^\dag_{i+nm}
\ee
and perform the explicit calculation.

We call 
\be
\mathbb{T}\,J_m(n,F)\,\mathbb{T}^{-1}=
\sum_i \mathfrak{F}(i)\mathfrak{F}(i+n)\mathfrak{F}(i+2n)\cdots\mathfrak{F}(i+n(m-1)) \psi_i\psi^\dag_{i+nm}
\ee
{\it canonical form} of $J_m(n,F)$.

{\bf Example 1.} Consider $F\equiv 1$. In this case $J_m(n,F)=J_{nm}$, where 
\be
J_k:=\sum_{i\in\mathbb{Z}} \psi_i\psi^\dag_{i+k}
\ee
is the current which is used to construct the KP tau function.

{\bf Example 2.} Consider $F(x) = x +N $, where $N$ is a positive number. 
Then
\be\label{J_m_n_F=x+N}
J_m(n,F)=\res_z \left(\left(z^n (D+N)\right)^m\cdot\psi(z)\right)\psi^\dag(z)\frac{dz}{z}=
\ee
\be
\sum_{i\in\mathbb{Z}} (N+i)(N+i+n)(N+i+2n)\cdots (N+i+n(m-1))\psi_i\psi^\dag_{i+nm}
\ee

In this case $J_1(n,F)=L_{n}+ N J_n$, where 
\be
L_n+NJ_n:=\sum_{i\in\mathbb{Z}} (i+N)\psi_i\psi^\dag_{i+n}
\ee
is the Virasoro generator. It's  canonical
form is
$$
\mathbb{T}^{-1}\,(L_n+NJ_n)\,\mathbb{T}=\sum_{i\neq -N} \psi_i\psi^\dag_{i+n}\,,
$$
which is different from $J_n$ because the term i=-N is absent in the sum in the right-hand side.

This abelian subalgebra (\ref{J_m_n_F=x+N}) with 
\be\label{n=-1}
n=-1
\ee
will be of use in matrix models below where
$N$ is the size of matrices.

\section{The model of normal matrices}

The model of normal matrices was introduced by O.Zaboronskii in \cite{Zabor}.

Consider the following model of normal matrices 
\be\label{norm}
I_N(\pb,\pb^*)=\int  e^{-\tr \left((M^\dag)^q M^p \right)}  
e^{\sum_{i>0} \frac 1i\left(p_i\tr M^i +p^*_i\tr (M^\dag)^i\right) } dM
\ee
\be
=C \int_{\mathbb{C}^N} \prod_{i=1}^N 
e^{-z_i^p \bar{z}_i^q+\sum_{i>0} \frac 1i\left(p_i z^i+p^*_i\bar{z}_i^i \right)}d^2 z_i
\ee
The case $p=q=1$ was intensively studied in the context of Laplacian growth problem 
\cite{MinZabrWigm}. The perturbation series of the integral (\ref{norm}) in the couplinf constants
$\pb$ and $\pb^*$ was considered in \cite{HO2003} and \cite{ShiotaOrl2004}.

We have
\be
I_N(\pb,\pb^*)=\sum_{\mu,\lambda}s_\mu(\pb)s_\lambda(\pb^*)\int s_\mu(\zb)s_\lambda({\bar \zb})
|\Delta(\zb)|^2\prod_{i=1}^N e^{-z_i^q{\bar z}_i^p}d^2 z_i
\ee

\subsection{Action on the vacuum and matrix models \label{MatrMod}}

The trivial example is
$$
e^{\sum_{m>0} \frac1m s_m J_{-m}}\cdot 1 = e^{\sum_{m>0} \frac1m  s_m p_m} .
$$

\begin{proposition}

For a given $n<0$ we obtain
\be
e^{\sum_{m>0} \frac1m s_m J_m(n,F)}\cdot 1=\sum_{\lambda} s_\lambda({\bf s})s_\lambda({\bf p})
\prod_{(i,j)\in\lambda} F(j-i)
\ee
where the sum is performed over the set of all partitions,
where $s_\lambda$ is the Schur function written as a polynomial of the KP higher times
where ${\bf s}$ is the set of higher times $(0,\dots,0,s_1,0,\dots,0,s_2,0,\dots)$ and 
$\tb=(t_1,t_2,t_3,\dots)$.
 
\end{proposition}

{\bf Example 1}

In particular, if $F(x)=x+N$ we get the commuting hierarchy which includes Virasoro element $L_{-1}$.

In this case
\be\label{main_KP_example}
e^{\sum_{m>0} \frac 1m s_m J_m(n,F)}\cdot 1 =\sum_{\lambda} s_\lambda({\bf s})s_\lambda({\bf t})
\prod_{(i,j)\in\lambda} (N+j-i)
\ee
The right hand side is the perturbation series for the famous two-matrix model
\be
\int e^{\tr XY + \sum_{m>1}\left(\frac 1m s_m \tr X^m + \frac 1m p_m\tr Y^m\right)} d\Omega(X,Y)
\ee
where $X$ and $Y$ are both $N\times N$ Hermitian matrices. The identical perturbation series one gets
in case $X$ and $Y$ are complex matrices and $Y=X^\dag$ and 
also in case 
 $X$ and $Y$ ar\begin{itemize}
  \item $X$, $Y$ both Hermitian
  \item $X=Y^\dag \in \mathbb{GL}_N(\mathbb{C})$  
  \item $X=Y^\dag$ normal (= diagonalizable via unitary matrix: $X=U\diag(x_1,\dots,x_N)U^\dag,\,U\in\mathbb{U}_N$)
 \end{itemize}

Let me recall that the famous one-matrix model can be obtained as the particular case of the two matrix
model (where both matrices are Hermitioan) via the specification of any of the sets (either ${\bf s}$ or $\tb$)
by putting all times to be zero except the second one. Say, is $s_m=\delta_{2,m}$ then by Gauss integration
over $X$ one obtains one-matrix model. Therefore with this speicification the series (\ref{main_KP_example})
serves also the one-matrix model; details see in \cite{HO2003}.

{\bf Example} ...$F=...$
\be\label{main_KP_example''}
e^{\sum_{m>0} \frac 1m s_m J_m(n,F)}\cdot 1 =\sum_{\lambda} s_\lambda({\bf s})s_\lambda({\bf t})
\prod_{(i,j)\in\lambda} (N+j-i)
\ee
\be
=\int e^{\tr X^\dag Y^\dag + \sum_{m>1}\left(\frac 1m s_m \tr X^m + \frac 1m p_m\tr Y^m\right)} d\Omega(X) d\Omega(Y)
\ee

 $X$ and $Y$ ar\begin{itemize}
  \item $X$, $Y$ both are unitary
  \item $X$, $Y$ are both complex
  \item $X$, $Y$ both are normal
 \end{itemize}e normal matrices and $Y=X^\dag$.

 \br\label{ensembles}
 For Hermitian matrices $X$ and $Y$ the measure is defined as
 \be\label{H-measure}
 d\Omega(X,Y)=C_N\prod_{i\ge j\ge N}d\Re X_{i,j} \prod_{i > j}d\Im X_{i,j} 
 \prod_{i\ge j\ge N}d\Re Y_{i,j} \prod_{i > j}d\Im Y_{i,j} 
 \ee
 For complex matrices $X$ the measure is defined as
 \be\label{C-measure}
 \left[e^{-\tr XX^\dag}\right] d\Omega(X,X^\dag) =C_N\left[e^{-\tr XX^\dag}\right]\prod_{i\ge j\ge N}d\Re X_{i,j} \prod_{i \ge j\ge N}d\Im X_{i,j}
 \ee
 where the Gaussian weight in square brackets can be included. However later we prefer to include the weight $\tau(XY)$ given by
 a KP tau function (details see below).

 For normal matrices $X=U\diag(x_1,\dots,x_N) U^{-1}$ 
 (where $x_1,\dots,x_N$ are the eigenvalues of $X$ and $U\in \mathbb{U}_N$)
 the measure is defined as
 \be
 d\Omega(X,X^\dag)=C_N\prod_{i<j\le N}|x_i-x_j|^2\prod_{1\le i \le N} e^{-|x_i|^2}d^2 x_i d_*U
 \ee
 where $d_*U$ is the Haar measure on $\mathbb{U}_N$.
 
 Here $N$ is the matrix size. Above it is supposed that, in each case, $C_N\int d\Omega =1$.
 \er

Let us mark that thank to the factor in the right hand side, actually, the sum is cutted if the length
of $\lambda$ exceeds $N$.

The canonical of this example is the sum 
\be\label{unitary_integral_series}
\mathbb{T}\,e^{\sum_{m>0} \frac 1m s_m J_m(n,F\equiv 1)}\mathbb{T}^{-1}\cdot 1=
\sum_{\lambda\atop \ell(\lambda)\le N} s_\lambda({\bf s})s_\lambda({\bf t})
\ee
(where $\mathbb{T}$ is the bosonic version of $\mathbb{T}$ above)
where the sum is restricted by partitions whose length do not exceed $N$. Such sums is the parturbation series
for Brezin-Gross-Witten (BGW) matrix model \cite{GrossWitten}
\be\label{BGW}
\int_{\mathbb{U}(N)} e^{\sum_{m>0} \frac 1m s_m \tr U^m + \frac 1m p_m\tr U^{-m}} d_*U
\ee
where $d_*U$ is the Haar measure on $\mathbb{U}_N$, we assume $\int_{\mathbb{U}_N}d_*U =1 $.
In different context the right hand side (\ref{unitary_integral_series}) was studied in the paper of Tracy and Widom \cite{TracyWidom}.

{\bf Example 3.} Consider $F(x)=(x+N)^{-1}$ for $x\neq-N$ and $F(x)=0$ for $x= -N$. Such $F$ has the same
characterisitc function as in the previous example. One obtains
$$
e^{\sum_{m>0} \frac 1m s_m J_m(n,F)}\cdot 1 = \sum_{\lambda\atop \ell(\lambda)\le N} s_\lambda({\bf s})s_\lambda(\tb)
\prod_{(i,j)\in\lambda}(N+j-i)^{-1}
$$
This series is equal to the value of the two-matrix integral where both matrices are unitary:
\be
\int_{\mathbb{U}_N\times \mathbb{U}_N} e^{\sum_{m>0}\frac 1m s_m\tr U^{m}_1} e^{\tr U^\dag_1U^\dag_2} e^{\sum_{m>0} \frac 1m p_m\tr U^{m}_2}
d_*U_1 d_*U_2
\ee

{\bf Example 4.} The rather general example is the perturbation series in the parameters $s_1,s_2,\dots$
and $t_1,t_2,\dots$ for the integral
\be
\int \tau_1({\bf s},X)\tau(XY)\tau_2(Y,\pb) d\Omega(X,Y) =\tau_3({\bf s},\pb)
\ee
where $\tau_{1,2}$ and $\tau$ are the following series
\be\label{tau_a}
\tau_a({\bf s},X)=\sum_{\lambda} s_\lambda(X)s_\lambda({\bf s})\prod_{(i,j)\in\lambda} f_a(j-i),\quad a=1,2
\ee
 and
 \be\label{interarction_tau}
\tau(XY) = \sum_{\lambda} s_\lambda(XY)s_\lambda(I_N)\prod_{(i,j)\in\lambda} g(j-i)
 \ee
 each of which is a KP tau function (more precisely: each is KP tau function of the hypergeometric type 
 \cite{KMMM}, \cite{OS}). Here $f_{1,2}$ and $g$ are functions on the lattice.
 
 In this case $\tau_3$ in the right hand side has the similar type (which was called hypergeometric type):
 \be
 \tau_3({\bf s},\tb)=\sum_{\lambda} s_\lambda({\bf s})s_\lambda(\tb)\prod_{(i,j)\in\lambda} F(N+j-i)\,,
 \ee
 where 
 $$
 F(x)=f_1(x)f_2(x)g(x)\kappa(x)\,,
 $$
 where the choice of $\kappa$ depend on the choice of the matrices; see Remark \ref{ensembles}:
 \begin{itemize}
  \item $X$, $Y$ both Hermitian
  \item $X=Y^\dag \in \mathbb{GL}_N(\mathbb{C})$  
  \item $X=Y^\dag$ normal 
  \item $X,Y\in\mathbb{U}_N$  
 \end{itemize}
 
This case includes the cases form previous examples. 
 
{\bf Example 5.} The genralization of (\ref{BGW}):
\be
\int_{\mathbb{U}_N} \tau_1({\bf s},U)\tau_2(U^{\dag},\pb)d_*U=
\sum_{\lambda} s_\lambda(X)s_\lambda({\bf s})\prod_{(i.j)\in\lambda} f_1(j-i)f_2(j-i)
\ee
 where the right  hand side  is of the similar type.
 
 Thus, the action of abelian groups of KP symmetries on the vacuum tau function (the tau function equal to  $1$)
 results in a number of matrix models.

 This was a collection of basically known facts related to matrix integrals and the KP hierarchy.

\label{lastpage}

\begin{thebibliography}{99}

   
\bibitem{GMMMO} A. Gerasimov, A. Marshakov, A. Mironov, A. Morozov, A. Orlov,
``Matrix models of two-dimensional gravity and Toda theory'', Nuclear Physics B 357 (2-3), 565-618 (1991); 
 S. Kharchev, A. Marshakov, A. Mironov, A. Orlov, A. Zabrodin,
``Matrix models among integrable theories: Forced hierarchies and operator formalism'',
Nuclear Physics B 366 (3), 569-601 (1991)
 
\bibitem{KacRadul} V.G. Kac and A. Radul, Quasi-finite highest weight modules over the Lie algebra of
differential operators on the circle, Comm. Math. Phys. 157 (1993), 429-457.

\bibitem{HO2003} J. Harnad and A. Yu. Orlov, ``Scalar products of symmetric functions and matrix integrals'', Theoretical and mathematical physics 137 (3), 1676-1690 (2003);
A. Yu. Orlov, Soliton theory, symmetric functions and matrix integrals
Acta Applicandae Mathematica 86 (2006) 131-158, arXiv:nlin/0207030  


\bibitem{Mac}  I.G. Macdonald,
{\sl Symmetric functions and Hall polynomials}, Second Edition, Oxford University Press,
1995


\bibitem{Mikhailov} A.V. Mikhailov, ``On the Integrability of two-dimensional Generalization of the 
Toda Lattice'', Letters in Journal of Experimental and Theoretical Physic
s, v.30, p. 443-448, 1979;
	A.V.Mikhailov, M. A. Olshanetski, A.M.Perelomov, Two-dimentional generalized Toda lattice, Comm.Math.Phys {\bf 79} (1981) no. 4 473-488
	

\bibitem{DJKM1} E. Date, M. Jimbo, M. Kashiwara and T. Miwa,
Physica {\bf 4D} (1982)  343-365

 \bibitem{JM}  M. Jimbo and T. Miwa,  
Publ. Res. Inst. Math. Sci. {\bf 19} (1983) 943-1001


\bibitem{DJKM2} E. Date, M. Jimbo, M. Kashiwara and T. Miwa, ``Transformation groups for soliton
equations'',  In: {\sl Nonlinear integrable systems - classical theory and quantum theory},
39-120. World Scientific (Singapore),  eds. M. Jimbo and T. Miwa (1983)

\bibitem{DJKM} E. Date, M. Jimbo, M. Kashiwara and T. Miwa, Publ. RIMS, Kyoto Univ. {\bf 18} (1982) 1077-1110

\bibitem{OS} A. Orlov, D.M. Shcherbin,
Theor.Math.Phys.
{\bf 128} (2001) 906-926

\bibitem{Orlov1988} A. Yu. Orlov,
``Vertex operator, ${\bar\partial}$-problem, symmetries, variational identities and Hamiltonian formalism for 2+1 D integrable systems''.
Nonlinear and Turbulent Processes in Physics 1987, ed. V. Baryakhtar., V.E.Zakharov —Singapore  (1988)

\bibitem{GrinOrl} P. G. Grinevich, A. Yu. Orlov, `` '' in ``Research Reports in Physics. Problems of Modern Quantum
Field Theory''
eds. A.A. Belavin, A. U. Klimyk, A. B. Zamolodchikov
Springer-Verlag Berlin. Heidelberg 1989 pp. 86-106
(republished in arXiv as arXiv-math-phys/9804019)

\bibitem{O-rational-soliton} A. Yu. Orlov, Hypergeometric functions as infinite-soliton tau functions,
Theoretical and mathematical physics 146 (2006) 183-206

\bibitem{PogrebkovSushko} A. K. Pogrebkov and V. N. Sushko, {\sl Quantization of the $(sin\psi)_2$ interaction in terms of fermion variables}, Translated from Teoreticheskaya i Mathematicheskaya Fizika,
{\bf 24} (1975) 425-429 (September, 1975). The original paper was submitted on May 15, 1975

\bibitem{UT} K.Ueno and K.Takasaki, \textit{Toda lattice hierarchy}, {\it Adv. Stud. Pure Math.} {\bf  4},  1-95 (1984).

\bibitem{GrossWitten} 
Mironov,A., Morozov,A. and Semenoff,G.: Unitary Matrix Integrals in the Framework of
the Generalized Kontsevich Model, Intern J Mod Phys A 11 (1996) 5031-5080

\bibitem{KMMM}  S. Kharchev, A. Marshakov, A. Mironov, A. Morozov,
\textit{Generalized Kazakov-Migdal-Kontsevich Model: group theory aspects},
International Journal of Mod Phys A10 (1995) p.2015


\bibitem{TracyWidom} C.Tracy, H.Widom, ``Random Unitary Matrices, Permutations and Painleve'',
Commun. Math. Phys. 207 (1999), 665-685
\bibitem{Dubr} B.A. Dubrovin, \textit{Symplectic field theory of a disk, quantum integrable systems, and Schur polynomials}, arxiv:1407.5824

\bibitem{E} Y. Eliashberg, \textit{Symplectic field theory and its applications,
Proceedings of the International Congress of Mathematicians}
Madrid, Spain, 2006.  2007 Europian Mathematical Society

\bibitem{ShiotaOrl2004}  A.Yu. Orlov and T. Shiota, ``Schur function expansion for normal matrix model and associated discrete matrix models'', Phys. Lett. A 343, 384-396 (2005)

\bibitem{Zabor} Chau, L-L. and Zaboronsky, O., ``On the Structure of Correlation Functions in the Normal Matrix Models'', Commun. Math. Phys. 196 (1998) 203-247

\bibitem{MinZabrWigm}
Mineev-Weinstein, M., Wiegmann, P. and Zabrodin, A., ``Integrable structure of interface
dynamics'', Phys. Rev. Lett. 84 (2000) 5106–5109

\bibitem{MMP} A. Mironov, A. Morozov, A. Popolitov, ``Commutative families in DIM algebra, integrable many-body systems and q,t matrix models''; arXiv:2406.16688 


\bibitem{LeznovSaveliev} 
A.N.Leznov, M.V.Saveliev, Communications in Mathematical Physics,
A. N. Leznov and M. V. Saveliev, Commun. Math. Phys. 89 (1983) 59.

\bibitem{MM} A. Mironov, A. Morozov, ``Spectral curves and $W$-representations 
of matrix models'',  J. High Energ. Phys. 2023 (2023) 116; arXiv:2210.09993 

\bibitem{MMMPWZ} A. Mironov, V. Mishnyakov, A. Morozov, A. Popolitov, Rui Wang, Wei-Zhong Zhao, ``Interpolating Matrix Models for WLZZ series'',  Eur. Phys. J. C 83 (2023) 377;
arXiv:2301.04107 

\bibitem{MM-1} A. Mironov, A. Morozov, ``Many-body integrable systems implied by WLZZ models'',
Physics Letters B842 (2023) 137964; arXiv:2303.05273

\bibitem{MMP} A. Mironov, A. Morozov, A. Popolitov, ``Commutative families in DIM algebra, integrable many-body systems and q,t matrix models'',
arXiv:2406.16688

\bibitem{DMP} Ya. Drachov, A. Mironov, A. Popolitov, ``$W_{1+\infty}$ and $\tilde{W}$ algebras, and Ward identities'', arXiv:2311.17738

\bibitem{MMNO} A.Mironov, A.Morozov, S.M.Natanzon and A.Yu.Orlov, ``Around spin Hurwitz numbers'',
 Lett.Math.Phys. 111 (2021) 124; arXiv:2012.09847 

\end{thebibliography}
\end{document}